\documentclass[conference]{IEEEtran}
\IEEEoverridecommandlockouts

\usepackage{mathtools}
\usepackage{amsmath,amsfonts, amsthm}
\usepackage{algorithmic}
\usepackage[linesnumbered,ruled,vlined]{algorithm2e}

\usepackage{textcomp}
\usepackage{booktabs}
\usepackage{xcolor, soul}
\usepackage{xspace}
\usepackage{hyperref}
\hypersetup{
    colorlinks,
    linkcolor={red!50!black},
    citecolor={blue!50!black},
    urlcolor={blue!80!black}
}

\usepackage{multirow}
\usepackage{enumitem}
\usepackage{graphicx}
\usepackage{textcomp}
\usepackage{url}

\usepackage[font=small,labelfont=bf]{caption}
\usepackage{subcaption}
\usepackage{cite}
\usepackage[numbers]{natbib}

\newcommand{\mat}[1]{\ensuremath{\mathbf{#1}}}

\newcommand{\nnz}{\textit{nnz}}

\newcommand{\mA}{\mathbf{A}} 
\newcommand{\mB}{\mathbf{B}}

\newcommand{\mC}{\mathbf{C}}

\graphicspath{{figures/}}

\begin{document}

\title{A sparsity-aware distributed-memory algorithm for sparse-sparse matrix multiplication}

\author{

 \IEEEauthorblockN{Yuxi Hong}
 \IEEEauthorblockA{\textit{Applied Math \& Computational Research Division} \\
 \textit{Lawrence Berkeley National Laboratory}\\
 Berkeley, USA \\
 YuxiHong@lbl.gov}
 \and
 \IEEEauthorblockN{Ayd{\i}n Bulu\c{c}}
 \IEEEauthorblockA{\textit{Applied Math \& Computational Research Division} \\
 \textit{Lawrence Berkeley National Laboratory}\\
 Berkeley, USA \\
 abuluc@lbl.gov}

}

\maketitle

\begin{abstract}
Multiplying two sparse matrices (SpGEMM) is a common computational primitive used in many areas including graph algorithms, bioinformatics, algebraic multigrid solvers, and randomized sketching. Distributed-memory parallel algorithms for SpGEMM have mainly focused on sparsity-oblivious approaches that use 2D and 3D partitioning. Sparsity-aware 1D algorithms can theoretically reduce communication by not fetching nonzeros of the sparse matrices that do not participate in the multiplication. 

Here, we present a distributed-memory 1D SpGEMM algorithm and implementation. It uses MPI RDMA operations to mitigate the cost of packing/unpacking submatrices for communication, and it uses a block fetching strategy to avoid excessive fine-grained messaging. Our results show that our 1D implementation outperforms state-of-the-art 2D and 3D implementations within CombBLAS for many configurations, inputs, and use cases, while remaining conceptually simpler.
\end{abstract}


\newcommand{\action}[1]{}
\newcommand{\new}[1]{}
\newcommand{\Aydin}[1]{}
\newcommand{\yuxi}[1]{}

\section{Introduction}
\label{sec:introduction}

Multiplying two sparse matrices (SpGEMM) is a widely utilized computational operation across various domains such as graph algorithms~\cite{kepner2011graph,Azad2015}, clustering~\cite{gilbert2006high,van2000graph}, bioinformatics~\cite{guidi2021bella,selvitopi2020distributed,besta2020communication,jain2019validating}, algebraic multigrid solvers~\cite{grey_comm_amg_2016}, and randomized sketching~\cite{sobczyk2022pylspack}. Distributed-memory parallel algorithms for SpGEMM have primarily emphasized sparsity-oblivious approaches involving 2D and 3D partitioning strategies~\cite{azad2021combinatorial,solomonik2015sparse}.

Sparsity-aware 1D algorithms have the potential to reduce communication theoretically by skipping the fetching of nonzeros from sparse matrices that are not involved in the multiplication. However, previous implementations have faced performance challenges (e.g., EpetraExt results~\cite{bulucc2012parallel} and 1D triangle counting implementation~\cite{Azad2015}), either due to the overhead of packing/unpacking submatrices or the increased number of messages for fine-grained messaging.

In this work, we introduce a distributed-memory 1D SpGEMM algorithm and its implementation. Our algorithm leverages MPI Remote Direct Memory Access (RDMA) operations to address the packing/unpacking overhead of submatrices during communication. Additionally, it employs blocking parameters to prevent excessive fine-grained messaging. Our algorithm offers several advantages over sparsity-oblivious 2D/3D algorithms: it eliminates the need for random permutation for load balancing, it utilizes both the inherent structure of the original sparse matrix and existing graph partitioners, and it integrates seamlessly with higher-level software using 1D partitioning, such as PETSc.

Our experimental results demonstrate that our 1D implementation outperforms state-of-the-art 2D and 3D implementations in CombBLAS~\cite{azad2021combinatorial} across various configurations, inputs, and use cases, all while maintaining conceptual simplicity.

\section{Background and Related Work}
\label{sec:background}
Given two sparse matrices $\mat{A} \in \mathbf{R}^{m \times k}$ and $\mat{B} \in \mathbf{R}^{k \times n}$, SpGEMM multiples \mat{A} and \mat{B} to get resulting matrix $\mat{C} \in \mathbf{R}^{m \times n}$, such that \mat{C} = \mat{A}\mat{B}.  Given sparse matrix \mat{A}, $nrows(\mat{A})$, $ncols(\mat{A})$, $nzc(\mat{A})$, $nnz(\mat{A})$ denote number of rows, column, non-zero columns, non-zero elements of \mat{A}.
The symbols used in this paper are listed in table \ref{tab:notation}.
\begin{table}[!t]
\centering
\caption{Notations used in this paper}
\begin{tabular}{ll}
\toprule
Symbol & Meaning \\
\toprule
 $P$  & total number of processes \\
 $p_i$  & $i^{th}$ processes in 1D column process grid layout \\
 $\mA \in \mathbf{R}^{m \times k}$ & first input matrix (distributed) \\
 $\mB \in \mathbf{R}^{k \times n}$ & second input matrix (distributed) \\
 $\mC \in \mathbf{R}^{m \times n}$ & output matrix (distributed) \\
 $\mat{A}_i$/$\mat{B}_i$/$\mat{C}_i$ & Part of $\mA$/$\mat{B}$/$\mat{C}$ in the $p_i$\\
 $\Vec{D}$ & global non-zero column id of $\mat{A}$ \\
 \multirow{2}{*}{$\Vec{H}_i$} & \multirow{2}{20em}{dense boolean vector of size $k$ that represents non-zero rows of $\mat{B}_i$} \\
 \\
\bottomrule
\end{tabular}
\vspace{-10pt}
\label{tab:notation}
\end{table}

We focus on distributed-memory SpGEMM algorithms. In our implementation, we use the Double Compressed Sparse Column (DCSC) data structure~\cite{hypersparse08} to store the local sub-matrices. We chose DCSC because its support in CombBLAS is more developed than that for CSC, but our algorithm would run on both with the same complexity bounds.
 For the local SpGEMM computation, we are using a hybrid version of Heap-based SpGEMM \cite{Azad2016} and Hash-based SpGEMM \cite{Nagasaka_SpGEMMPerfOpt_2019}. 

\subsection{Distributed Parallel SpGEMM Algorithms}
Many block partition algorithms have been proposed for distributed-memory SpGEMM workload. They can be categorized into 1D, 2D and 3D algorithms according to how the computational cube is partitioned to processes and which matrices move~\cite{ballard_communication_2013}. 

Bulu\c{c} et al. \cite{Buluc2008icpp} proposed 2D sparse SUMMA algorithm which is based on dense SUMMA algorithm \cite{Geijn1995}. The $P$ processes are organized in a square $\sqrt{P} \times \sqrt{P}$ 2D process grid. Square process grid is not necessary and is only for simplicity here. The communication pattern ignores the sparsity structure if random permutation is performed. Azad et al. \cite{Azad2015} extends this setting to $\sqrt{\frac{P}{c}} \times \sqrt{\frac{P}{c}} \times c$ 3D process grid layout. $c$ is the number of layers, the size of third dimension. The conclusion is that proposed Split-3D-SpGEMM algorithm has better performance over Sparse 2D SUMMA algorithm due to communication volume reduction.

Ballard et al.~\cite{ballard_communication_2013} analyzed the communication cost of 1D, 2D, 3D SpGEMM algorithms on Erd\H{o}s-R\'enyi random matrices. Two algorithms in 1D partitioning are mentioned: naive block row algorithm and improved block row algorithm. 
The naive block row algorithm computes \mat{C} by leaving \mat{A} and \mat{C} to be stationary and exchanging \mat{B} in a ring style among all processes. This will introduce huge communication volume since each process will receive a copy of \mat{B} for local computation. 
The improved block row algorithm exchanges the needed rows of \mat{B} that is only needed for the local computation. 

The analysis by Ballard et al.\, shows the improved block row 1D SpGEMM algorithm communicates asymptotically fewer bytes of data (as the number of processors increases) compared to the 2D SpGEMM algorithms when dealing with sufficiently sparse Erd\H{o}s-R\'enyi matrices. In particular, the improved block row 1D SpGEMM algorithm communicates a multiplicative factor of $O(\sqrt{P}/d)$ fewer words than the 2D SpGEMM algorithms, where $P$ is the number of processors and $d$ is the expected number of nonzeros in each row or column of an Erd\H{o}s-R\'enyi matrix. The communication volume of the improved block row 1D SpGEMM algorithm also matches communication volume of the 3D SpGEMM algorithms (up to logarithmic factors), while being more flexible and simpler.

Our idea is similar to the improved block row algorithm, however we use RDMA to remove the ring style exchange of \mat{B}. We identify that random graphs are the worst use cases for 1D SpGEMM and better partition methods can be used onto the input matrices to further minimize the communication volume.

\subsection{Random Permutation and Graph Partitioning}
Rearranging the rows and columns of a sparse matrix is a common technique in sparse computation to achieve load balancing, gather non-zero elements for coalesced memory access, and minimize communication volume in distributed algorithms. There are many techniques to achieve this goal. In this paper, we focus on random permutation, graph and hypergraph partitioning.

\subsubsection{Random Permutation}

\textbf{Random permutation} is a common and effective preprocessing step for SpGEMM to achieve load balancing. It has been used in many distributed SpGEMM algorithms \cite{Azad2016,azad_hipmcl_2018,hussain-batch3dspgemm-2021}. This technique applies a symmetric permutation to the matrices, which equivalently means randomly relabeling the vertices of the graph that corresponds to the sparse matrix. Mathematically, instead of computing $\mat{C}=\mat{A}\mat{B}$, it computes $\mat{P}\mat{C}\mat{P}^T=(\mat{P}\mat{A}\mat{P}^T)(\mat{P}\mat{B}\mat{P}^T)$, where \mat{P} is a permutation matrix. The downside of the random permutation is that it disrupts the sparsity pattern of the original sparse matrix and the data movement volume could be very high if the algorithm is not properly designed. The permutation also introduces extra computation and communication. In 2D and 3D sparse SUMMA algorithms, the input matrices must be sent and received through the network multiple times. In this paper, we will show that in many use cases, sparsity-aware 1D SpGEMM algorithms can utilize the original sparsity pattern, making random permutation unnecessary.

\subsubsection{Graph Partitioning}
\textbf{METIS} and \textbf{ParMETIS} \cite{Karypis1995,parmetis1996} are software tools designed for partitioning unstructured graphs, finite element meshes, and for producing fill-reducing orderings of sparse matrices. 
METIS is a serial program that provides high-quality graph partitioning and sparse matrix ordering solutions. It utilizes multilevel algorithms that reduce the size of the graph by collapsing vertices and edges, partitioning the smaller graph, and then uncoarsening to obtain a partition for the original graph. This technique allows METIS to efficiently partition large-scale graphs, which is crucial in reducing communication in distributed computations.
ParMETIS extends METIS to distributed memory environments, enabling it to handle very large graphs that do not fit into the memory of a single machine. It parallelizes the multilevel graph partitioning process, making it suitable for distributed computing environments where the graph data is distributed across multiple processors. ParMETIS is widely used in parallel computing applications where the efficient division of computational work among processors is essential to minimize communication overhead and balance workloads.
Both METIS and ParMETIS are integral tools in high-performance computing, particularly in fields like scientific computing, where efficient graph partitioning can significantly enhance the performance of parallel algorithms and simulations.

\textbf{Hypergraph partitioning} is a technique used to divide a hypergraph into smaller, balanced parts while minimizing the connections (or "cut") between these parts. Unlike traditional graph partitioning, which involves dividing a graph where each edge connects exactly two vertices, hypergraph partitioning deals with hypergraphs, where an edge (called a hyperedge) can connect more than two vertices. This makes hypergraph partitioning particularly suited for modeling complex relationships and dependencies that cannot be easily captured by simple graphs.
Akbudak et al. \cite{Akbudak2013,Akbudak2014,Akbudak2018} proposed several hypergraph models and bipartite graph models for 1D SpGEMM. On top of these models, it uses hypergraph partitioner to achieve simultaneous partitioning of both input/output matrices and to reduce the communication cost.
Graph partitioning is also used in other computational primitives like SpMM \cite{Acer2016,gianinazzi2024arrow}.

\subsection{SpGEMM Applications}
SpGEMM has numerous applications in scientific computing and data analysis, but a comprehensive review of all SpGEMM applications is beyond the scope of our paper. Here we will describe the squaring operation and other two important applications that we will use in our benchmarks. 

\subsubsection{Squaring}
\textbf{Squaring} a sparse matrix \mat{A} in a distributed setting involves multiplying it by itself. Though it is not an application itself, it powers a lot of applications such as MCL \cite{azad_hipmcl_2018,Dongen2008}, and still remains to be the bottleneck. The squaring operation is found in various other applications, such as graph algorithms, sparse linear algebra, and machine learning. The challenge lies in efficiently handling the sparsity of the matrix, which leads to irregular data access patterns and communication overhead in distributed systems. Each processor works on a portion of the matrix, performing local computations and then exchanging data with other processors to complete the multiplication. The irregular distribution of non-zero elements can cause load imbalance, while the communication between processors, often the most time-consuming aspect, can significantly impact performance. Memory management is also complex; the product matrix \mat{C} may have more non-zero elements than \mat{A}, requiring careful prediction and allocation of resources.

\subsubsection{Algebraic Multigrid Solvers}
\textbf{Algebraic Multigrid (AMG) solvers} are advanced iterative methods used to efficiently solve large, sparse linear systems of equations, which commonly arise from discretized partial differential equations (PDEs) in scientific computing. Unlike geometric multigrid methods, which rely on the underlying geometric structure of the problem, AMG solvers operate purely on the algebraic structure of the matrix. They work by constructing a hierarchy of progressively coarser approximations of the original problem, where the solution is iteratively refined. The key advantage of AMG solvers is their scalability and ability to handle complex, unstructured grids without requiring explicit geometric information. This makes them particularly effective for solving problems in areas such as computational fluid dynamics, structural analysis, and other large-scale simulations where traditional solvers may be inefficient or impractical.

The Galerkin product ($\mat{R}\mat{A}\mat{P}$) has been a performance bottleneck of Algebraic Multigrid (AMG) \cite{Bell2012,grey_comm_amg_2016}. The Galerkin product is used in the restriction/prolongation steps of AMG to go from fine to coarse grids.
The interpolation operator \mat{P} is the transpose of the restriction operator \mat{R}. In other words, the Galerkin product can be written as $\mat{R}^T\mat{A}\mat{R}$.
\mat{R} is usually short and fat, so \mat{P} is usually tall and skinny.
In order to compute restriction operator \mat{R}, Bell et al. \cite{Bell2012} uses distance-2 Maximal Independent Set (MIS-2). It is a generalization of the MIS where no two vertices are distance-2 neighbors, meaning that those two vertices do not have a common neighbor. Azad et al. \cite{Azad2016} also utilizes MIS-2 to compute the restriction operator. Two SpGEMMs are used to compute the Galerkin product. We refer to $\mat{R}^T\mat{A}$ as left multiplication and $(\mat{R}^T\mat{A})\mat{R}$ as right multiplication. For large problem sizes, these two SpGEMMs can account for up to 80\% of the total construction time. Moreover, the Galerkin product will be used in multiple iterations, making it essential to optimize its SpGEMM operations.

\subsubsection{Betweenness Centrality}
\textbf{Betweenness Centrality (BC)} is a measure used in network analysis to quantify the importance of a node within a network based on the number of shortest paths that pass through it. Specifically, it calculates the extent to which a node lies on paths between other nodes, serving as a bridge or mediator in the flow of information or resources across the network. A node with high betweenness centrality has significant influence over the interactions between other nodes because it often connects otherwise disparate parts of the network. This metric is widely used in various fields, such as sociology, biology, and computer science, to identify key nodes that facilitate communication or control the flow within networks, such as influential individuals in social networks or critical routers in communication networks.

The algorithm uses the number of shortest paths going through a vertex as indicator of the centrality. The definition of centrality $g(v)$ of a vertex $v$ is given in Eq.\ref{eq:bc}, $\sigma_{st}$ is shortest path from vertex $s$ to vertex $t$.
\begin{equation} \label{eq:bc}
    g(v)=\sum_{s\neq v\neq t}{\frac{\sigma_{st}(v)}{\sigma_{st}}}
\end{equation}
Breadth-first search (BFS) is the most commonly used algorithm to compute the shortest path \cite{Brandes2001,david-bader-bc-2006,david-bader-fasterbc-2013,tan-parallelbc-2009} on unweighted graphs. Solomonik et al. \cite{solomonik2017scaling} uses the Bellman-Ford shortest path algorithm to compute the shortest distances and develops Maximal-Frontier Betweenness Centrality (MFBC) algorithm.
BC on unweighted graphs was also one of the first algorithms implemented in the original CombBLAS~\cite{bulucc2011combinatorial}.

In this paper, we focus on a batched approximate version of the Brandes Algorithm \cite{Brandes2001}. At each iteration, the algorithm computes single-source shortest paths from a source vertex, followed by a backward graph traversal for tallying betweenness centrality contributions of each shortest path to the vertices along those shortest paths. The exact algorithm iterates over all vertices whereas the approximate algorithm iterates over randomly chosen $K$ source vertices. The batched algorithm splits $K$ source vertices into multiple batches. For each batch, we will use multi-source BFS (SpGEMM kernel inside) as forward search method. In the same batch, backward sweep (SpGEMM kernel inside) is used to compute the scores of each vertex. The most time consuming operations in BC is forward search and backward sweep. Thus here we only present the results of forward search and backward sweep.

\begin{figure}[t]
    \centering
    \includegraphics[width=\linewidth]{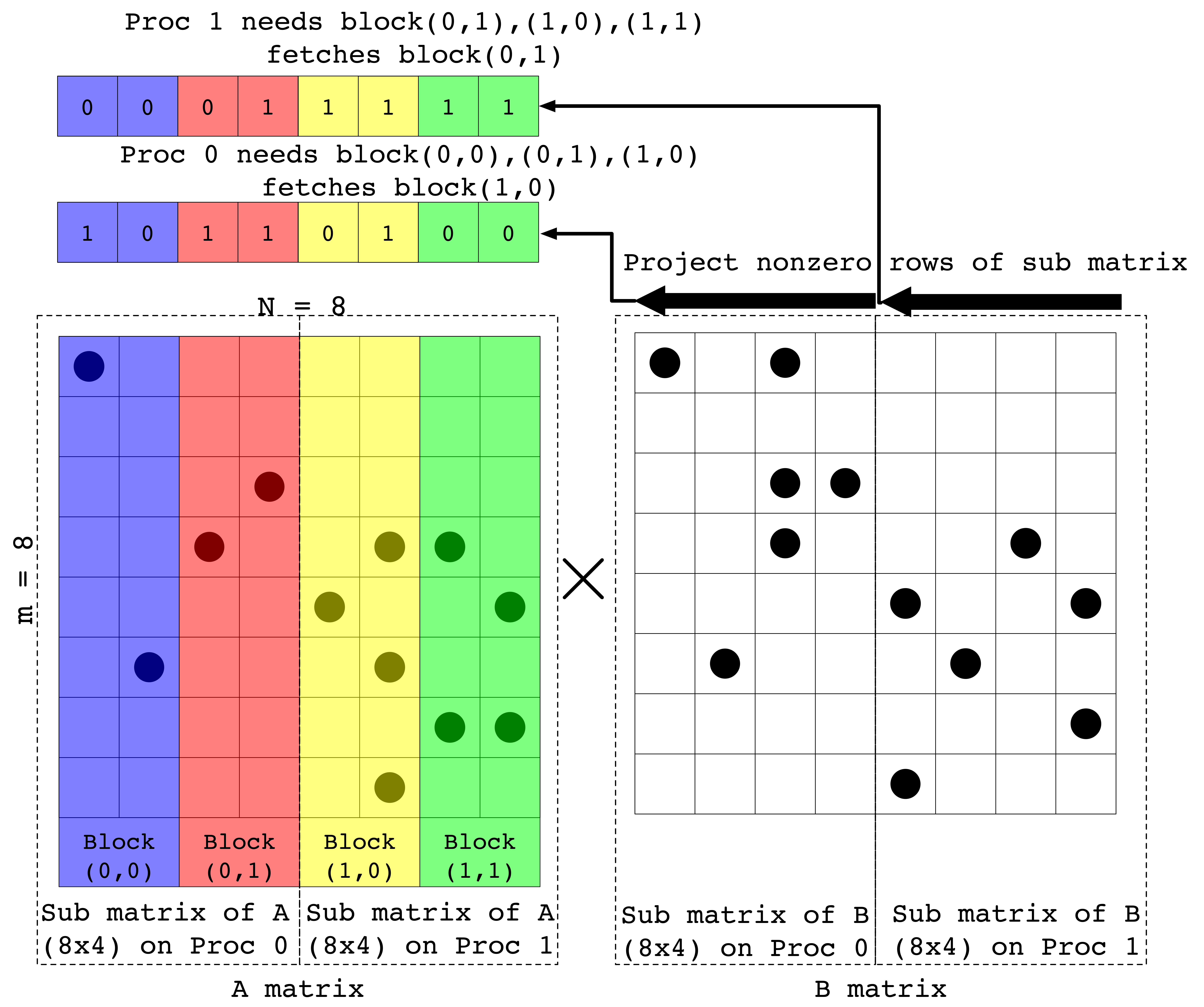}
    \caption{An example of sparsity-aware 1D SpGEMM algorithm (Algorithm~\ref{alg:spgemm1d-cbc}) and block fetching strategy (Algorithm~\ref{alg:kchunkgroup}).}
    \label{fig:spgemm1dexplain}
\end{figure}

\section{Sparsity-aware 1D SpGEMM Algorithm}
\label{sec:spgemm1d}

\subsection{Main Algorithm and Block Fetch Strategy}

We will describe our algorithm by using an example. Figure~\ref{fig:spgemm1dexplain} describes the sparsity-aware 1D SpGEMM algorithm on two processes. The matrix size of \mat{A} and \mat{B} are both $8 \times 8$. 
Two processes $p_0$ and $p_1$ are arranged in a 1D column layout. Each process owns one submatrix (size of the submatrix is $8 \times 4$) of two input matrices and the output \mat{C}. We are using column-by-column formula to compute the SpGEMM. 

In our sparsity-aware 1D SpGEMM algorithm, \mat{B} and \mat{C} is stationary, communication is only necessary for fetching remote data of \mat{A}. Each  process projects the non-zero rows of its submatrix of \mat{B} to $\Vec{H}_i$ of length 8. 
We partition the submatrix of \mat{A} on each process into 2 blocks, so there is 4 blocks in total. Each process will fetch a block as long as the block is non-empty and the corresponding sub vector of $\Vec{H}_i$ has at least 1 non-zero row. For example, non-zero rows of \mat{B} of $p_0$ is $\Vec{H}_0 = [1, 0, 1, 1, 0, 1, 0, 0]$. The subvector $\Vec{H}_0(4:7) = [0, 1, 0, 0]$ indicates that $p_0$ needs the second column of $\mat{A}_1$ from $p_1$. 

Ideally we would like to only fetch the necessary columns from $\mat{A}_1$. But this increases the RDMA message numbers when the required columns are not contiguous. We will use the \emph{block fetch strategy} to significantly reduce the RDMA messages number. In this example, each submatrix $\mat{A}_i$ are split into 2 blocks. Therefore, we will fetch block(1,0) from $p_1$, even though the first column of $\mat{A}_1$ is not needed in the local computation of $p_0$. 

After $p_i$ gets the remote columns of \mat{A} that are needed for computing $\mat{C}_i$, it can launch the local SpGEMM individually and generate the $\mat{C}_i$. No communication of output matrix is needed since $\mat{C}$ is naturally in distributed 1D layout after the computation.

\begin{algorithm}[t]
\caption{Sparsity-aware 1D SpGEMM Algorithm}
\label{alg:spgemm1d-cbc}
\DontPrintSemicolon
\KwData{Distributed sparse matrix $\mat{A}^{m \times k}$ and $\mat{B}^{k \times n}$. They are both split into a 1D process grid along the column. Each process $p_i$ owns a slice of \mat{A} and \mat{B}: $\mat{A}_i^{m \times k_i}$ and $\mat{B}_i^{k \times n_i}$, such that $\sum k_i = k, \sum n_i = n$. There are $P$ processes in all.}
\KwResult{Distributed sparse matrix $\mat{C}^{m \times n} = \mat{A} \times \mat{B}$. $\mat{C}$ is also split into 1D process grid along the column. Each process owns $\mat{C}_i^{m \times n_i}$ and $\sum n_i = n$.}

Create two MPI Windows for row id and numeric values of \mat{A}. \;

\textbf{Allgather} non-zero columns id of \mat{A} ($\Vec{D}$) 
and nnz in each column. \;

\For{Each process $p_i$ in parallel}{
Store non-zero rows id of $\mat{B}_i$ in $\Vec{H}_i$ \;

Create the required columns id of \mat{A}: $\Vec{\tilde{D}}$ = $\Vec{H}_i \cap \Vec{D}$ \;

Generate column blocks that are needed to be fetched from remote MPI processes using block fetch strategy described in Algorithm~\ref{alg:kchunkgroup}\;

Use passive-target RDMA Calls (\textbf{MPI\_Get}) to fetch the remote column block data (row id and numeric values) of \mat{A} from two MPI Windows. \;

Create a new sparse matrix $\tilde{\mat{A}}$ which only contains data needed for local results computation. \;

Compute local results $\mat{C}_i = \tilde{\mat{A}} \times \mat{B}_i$ \;

}

\end{algorithm}

A more thorough explanation of the sparsity-aware 1D SpGEMM algorithm is in Algorithm~\ref{alg:spgemm1d-cbc}. MPI processes are arranged in 1D column layout, with first process in the left most side and last process in the right most side. All MPI processes expose their local sub-matrix $\mat{A}_i$ to others. All MPI processes own a copy of non-zero column id and prefix sum of non-zero elements in the column. They are used to calculate the memory space for remote data and indexing of the RDMA calls. Everyone fetch the data using \textbf{MPI\_Get} and compute the local results $\mat{C}_i$ independently. There is no overlap between communication and computation. Note that we create a new sparse matrix  $\tilde{\mat{A}}$ for the local computation. If there is sparsity pattern in the $\mat{A}$, the $\tilde{\mat{A}}$ will be much smaller than $\mat{A}$. This will also improve the data locality of local SpGEMM compare to the method that use \mat{A} to compute results.

The block fetch strategy is described in Algorithm~\ref{alg:kchunkgroup}. Since the non-zero columns of $\mat{A}_i$ might be very large. It will be inefficient to use 1 RDMA call per column. 
\new{
However, it is inefficient to exchange all the local columns of \mat{A} to other MPI processes.
} \Aydin{I didn't understand this sentence, why is it impossible, and what is ``needed data''}. Our solution is to chunk the remote $\mat{A}_i$ matrix into $K$ blocks, as long as we need any column in the block, we will fetch all the data inside the block. By doing block, we keep the limit of RDMA messages to $K$ for each remote process.

\begin{algorithm}[t]
\caption{Block Fetching Algorithm}
\label{alg:kchunkgroup}
\DontPrintSemicolon
\KwData{$K$: non-zero column split number (e.g. 2048). $\Vec{\tilde{D}}$: required column id vector. $\Vec{H}$: boolean hit vector. $\Vec{\tilde{D}}$ and $\Vec{H}$ are of the same length. $\Vec{H}_i = 1$ means column $\Vec{\tilde{D}_i}$ is needed in the local computation. Thus the corresponding data is also needed to be fetched using RDMA.}
\KwResult{$M$: Total RDMA calls after grouping, $M \leq K$. $\Vec{R}$: $M$ ordered pairs. The start and end column id of each pair are associated with the RDMA calls. Union of the intervals in the $\Vec{R}$ must cover the required column id $\Vec{\tilde{D}}$.}

M = 0 \;
split the ordered non-zero column id into K groups $G$. \;

\For{each group $G_i$}{
choose = false \;
\For{each column $c_j$ in the $G_i$}{
    \If{corresponding $H_j$ = 1}{
        choose = true \;
        break \;
    }
}
\If{choose}{
    M += 1 \;
    add begin and end column of $G_i$ as a pair into $\Vec{R}$  \;
    }
}

\end{algorithm}


\subsection{Graph Partitioning}
\new{For symmetric sparse matrices, if the original sparse structure is not well presented, we can further use graph partition methods to split the sparse matrices.}

As stated in the Section~\ref{sec:background}, random permutation is usually a good preprocessing strategy for sparsity-oblivious 2D and 3D distributed-memory SpGEMM algorithms. However, it is  not a good strategy for our sparsity-aware 1D SpGEMM algorithm. This is because our algorithm communicates less when nonzeros are clustered as opposed to scattered evenly. Since \mat{B} and \mat{C} are stationary in the algorithm, we need to to minimize the communication volume of \mat{A}. In other words, we need to minimize the non-zero rows of $\mat{B}_i$ that falls in the remote processes. We prefer preserving the sparsity structure of \mat{A} to randomly permuting the matrix. 

Keeping the original sparsity structure of the matrix intact can however introduce load imbalance, especially in communication volume. Also, in the case that non-zero elements are not naturally clustered in the sparse matrix, the original sparsity structure can incur high communication volume. Graph partitioning can be an alternative way to improve the performance. 

We assign a weight to each vertex for balancing the amount of sparse flops (the nontrivial scalar multiplications of the form $a_{ij} \cdot b_{ij}$ where both $a_{ij}\neq0$ and $b_{ij}\neq 0$) each process needs to perform. The weight value is the square of non-zero elements of the column. Alternatively when considering the graph view of the sparse matrix, the weight is the square of the degree of the vertex corresponding to that column. The reason is that we not only need to partition the sparse matrix evenly into $P$ parts but also we need partition the work (\emph{sparse flops}) each local SpGEMM will perform evenly to each process. 

Using the outer-product view of SpGEMM, it is known that~\cite[Proof of Th 13.1] {bulucc2011implementing}, \cite[Eq 3.5]{ Akbudak2014} the \emph{sparse flops} of multiplying two sparse matrices (e.g. \mat{A} and \mat{B}) is inner-product of the column non-zero counts of \mat{A} and row non-zero counts of \mat{B}. Since METIS requires the graph to be undirected, the sparse matrix itself should be symmetric\Aydin{I still don't understand what we do when the input is not symmetric. We don't use METIS? We should explain what we do in that case clearly}. When squaring a symmetric sparse matrix, the column non-zero elements of \mat{A} equals row non-zero elements of \mat{B}. Then the flops assigned to each vertex will be the square of the corresponding column non-zero elements. Though the flops count is calculated through squaring operation, we use it as an approximate flops count for the restriction operator as well as for betweenness centrality.

\new{As stated in the beginning of the section, we can only use graph partition when the sparse matrix is symmetric, as we are restricted by METIS input criteria. For asymmetric sparse matrix, we only use its original sparse structure.}

\begin{figure}[t]
    \begin{minipage}{.48\linewidth}
    \includegraphics[width=\linewidth]{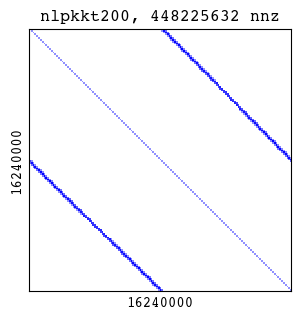}
    \caption{nlpkkt200 visualization}
    \label{fig:ss_nlpkkt200}    
    \end{minipage}
    \begin{minipage}{.48\linewidth}
    \includegraphics[width=\linewidth]{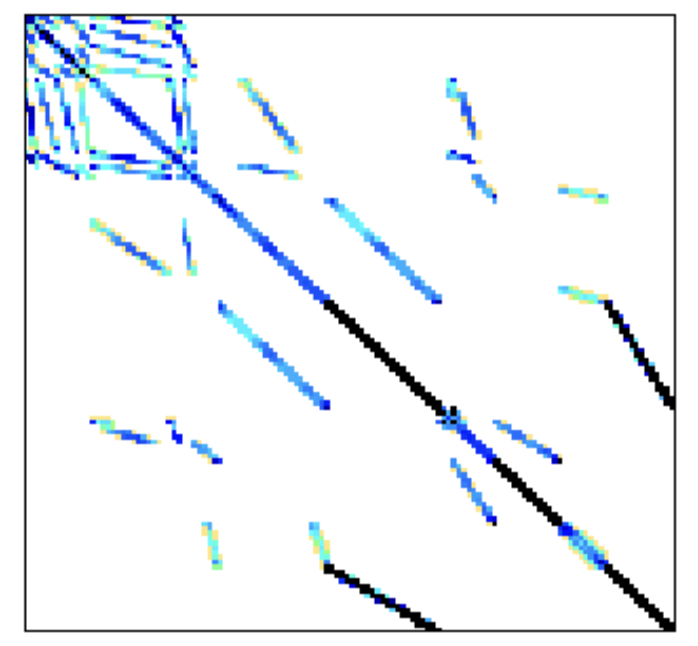}
    \caption{hv15r visualization}
    \label{fig:ss_hv15r}    
    \end{minipage}
\end{figure}

\subsection{Implementation Details}
We implement our algorithm in CombBLAS library. The Programming model is MPI + OpenMP. 
We follow the tradition of CombBLAS that number of MPI processes is a perfect square. We also conduct an exhaustive search to find out the best MPI processes and OpenMP threads configuration for sparsity-aware 1D SpGEMM algorithm.
We use ParMETIS, a parallel implementation of METIS, to partition the input graph if necessary. The ParMETIS uses 64 bits width index and double precision for the numeric values.

For the right multiplication of Galerkin product, which is the $(\mat{R}^T\mat{A})\mat{R}$ part, we implement an distributed-memory outer-product 1D SpGEMM algorithm. Ballard et al.~\cite{grey_comm_amg_2016} previously showed that the outer-product SpGEMM algorithm is the best 1D algorithm for the right multiplication. 
\begin{algorithm}[ht]
\caption{Outer-Product 1D SpGEMM Algorithm}
\label{alg:spgemm1d-op}
\DontPrintSemicolon
\KwData{Same input as Algorithm~\ref{alg:spgemm1d-cbc}}
\KwResult{\mat{C}=\mat{A}\mat{B}}
Redistribute \mat{B} so that $p_i$ owns $i^{th}$ row block \;
Launch local SpGEMM to get part of Outer-Product result \mat{C} \;
Redistribute Outer-Product result \mat{C} and merge results \;
\end{algorithm}

\section{Experiment Results}
\label{sec:results}

\begin{figure}[t]
    \centering
    \begin{minipage}{\linewidth}
    \includegraphics[width=\linewidth]{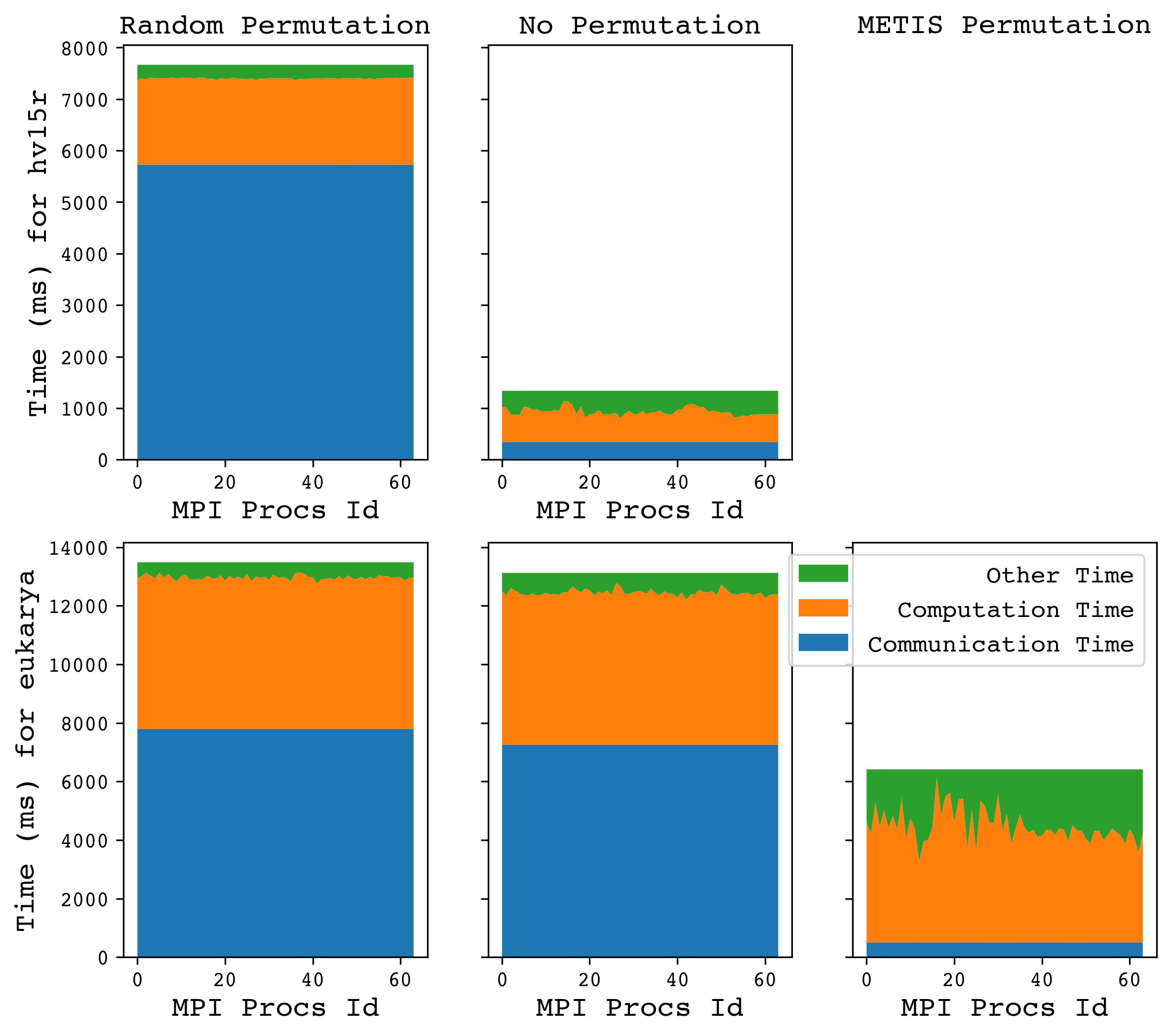}
    \caption{Impact of permutation strategies on hv15r (upper two figures) and eukarya (lower three figures) datasets in squaring operation. Note that hv15r doesn't have METIS Permutation.}
    \label{fig:AA_permimpact_timebreakdown_eukhv15r}
    \end{minipage}
\end{figure}

We evaluate the performance of our algorithms on NERSC Perlmutter system \cite{nersc_perlmutter_architecture}, a HPE Cray EX supercomputer. Each CPU node on Perlmutter system has 2 AMD EPYC 7763 CPUs (codename Milan), with 64 cores per CPU and 512 GB of DDR4 memory in total. Each CPU node is connected to 1 HPE Slingshot 11 network interface controller (NIC) in a 3-hop dragonfly network topology. We rely on default Cray MPICH MPI implementation and GNU Programming Environment on Perlmutter system.


\begin{table}[ht]
    \centering
    \caption{Statistics information of sparse matrices used in our experiments.
    M denotes million.}
    \begin{tabular}{l r r r r r}
    \toprule
    Matrix ($\mA$) & rows & columns & $\nnz(\mA)$ & symmetric \\
    \midrule
    queen\_4147 & 4M & 4M & 330M & Yes\\
    stokes & 11M & 11M & 350M & No\\
    eukarya & 3M & 3M & 360M & Yes\\
    hv15r & 2M & 2M & 283M & No\\
    nlpkkt200 & 16M & 16M & 448M & Yes\\
    \bottomrule
    \end{tabular}    
    \label{tab:dataset}
\end{table}

We use real-world sparse matrices to benchmark our proposed algorithm. Table~\ref{tab:dataset} describes sparse matrices used in our experiments. Figure~\ref{fig:ss_nlpkkt200} and Figure~\ref{fig:ss_hv15r} visualize the nlpkkt200 and hv15r dataset. The non-zero elements are clustered together in some matrices. However, they are not simple enough to categorize as banded or diagonal block matrices. 

\subsection{Squaring}
\begin{figure}[t]
    \centering
    \begin{minipage}{\linewidth}
    \includegraphics[width=\linewidth]{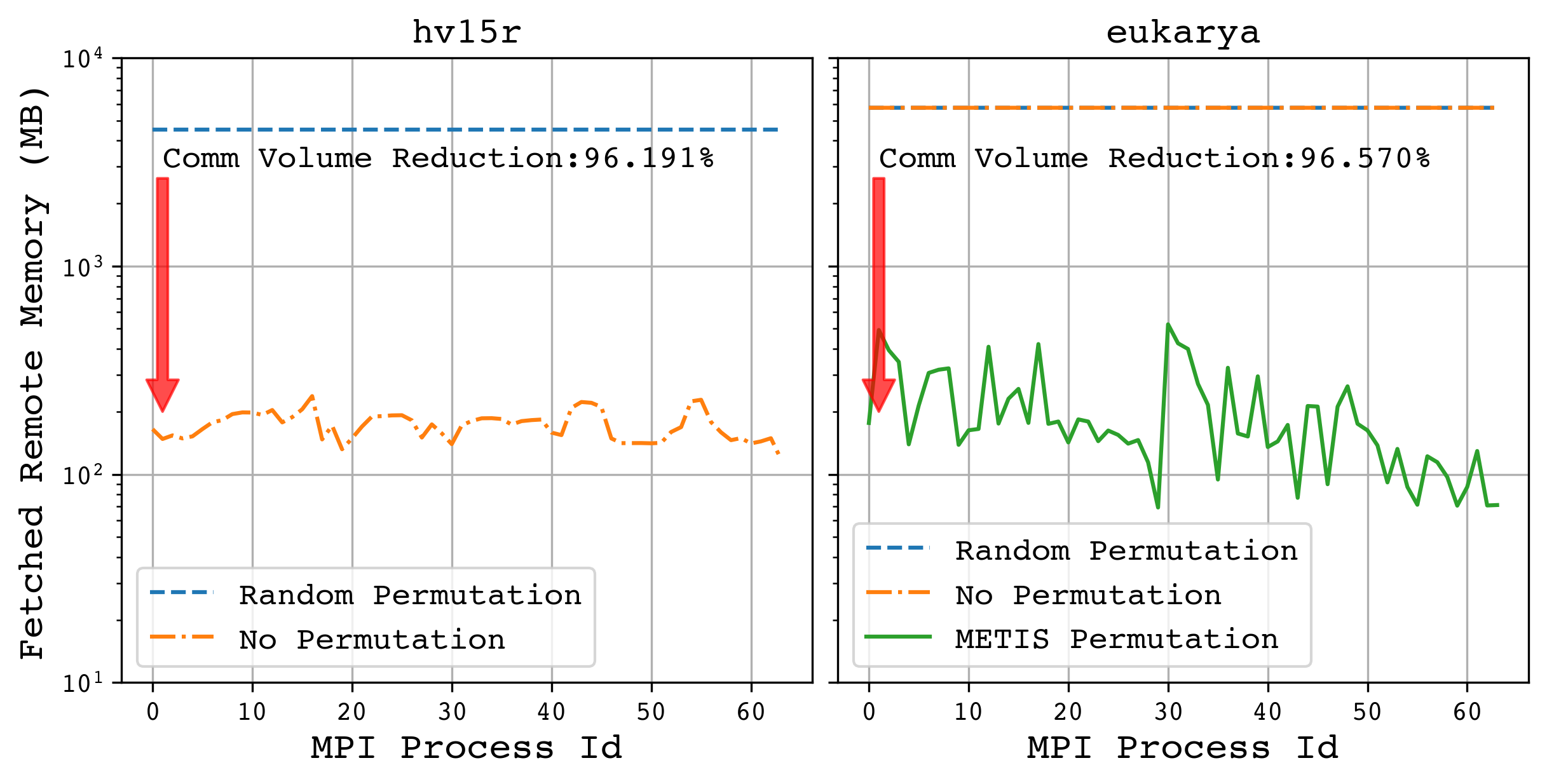}
    \caption{Communication volume comparison of different permutation strategies in squaring operation.}
    \label{fig:eukhv15r_commana_AA}    
    \end{minipage}
    \begin{minipage}{\linewidth}
    \includegraphics[width=\linewidth]{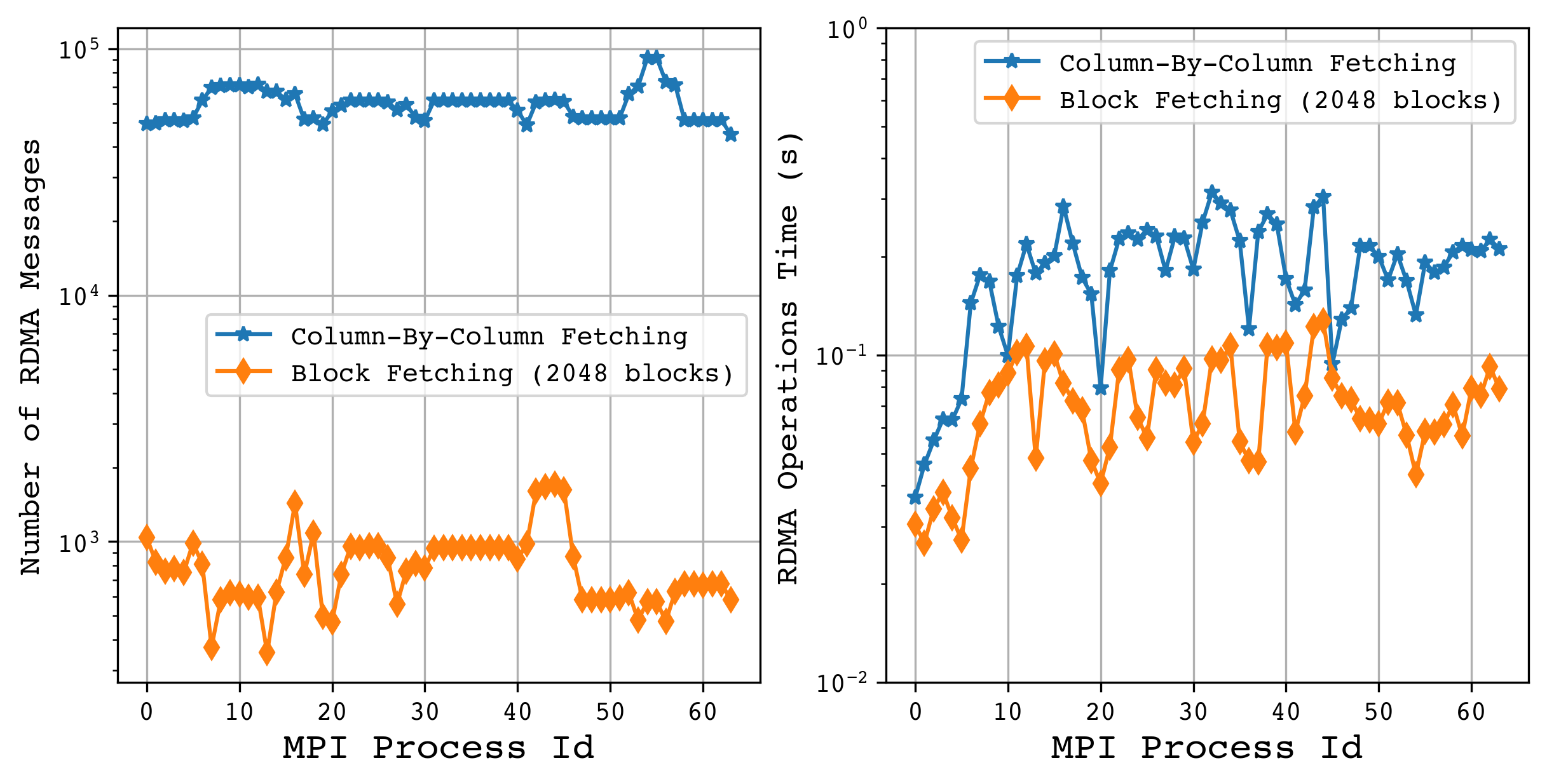}
    \caption{Block fetch strategy analysis on hv15r dataset in squaring operation.}
    \label{fig:hv15r_commana_AA}    
    \end{minipage}
\end{figure}
We first present the results of squaring a sparse matrix. Here \mat{A} = \mat{B}, and $m=k=n$. We use hv15r and eukarya as showcases to illustrate the permutation impact, RDMA communication analysis and block fetch algorithm. 
\subsubsection{Permutation Impact to Performance}
Figure~\ref{fig:AA_permimpact_timebreakdown_eukhv15r} shows the time breakdown of sparsity-aware 1D SpGEMM algorithm on 4 nodes with 64 MPI processes and 16 OpenMP Threads. We are showing per MPI process times in each figure in order to highlight any potential load imbalance issues. We are reporting communication time, computation time and other time. Communication time includes RDMA requests that fetch the remote \mat{A} data. Computation time includes the local SpGEMM computation. Other time refers to the creation and deletion of auxiliary arrays and data structures, such as the creation time of the local DCSC format object. Other time also includes time spent exchanging the non-zero columns and non-zero elements of $\mat{A}_i$. For both input matrices shown, random permutation is the worst performer because the sparsity structure is removed after applying a random symmetric permutation. Compared to the randomly permuted case, using the original sparsity structure of the matrix reduces the communication time in hv15r by $16.86\times$ (from 5725.5 ms to 339.4 ms), which translates into a $5.73\times$ speedup. As a side outcome, the computation time is also shortened. This is because we only compact what we need to compute the $\mat{C}_i$ into the new sparse matrix $\tilde{\mat{A}}$ matrix and achieve a better data locality. 

In the eukarya dataset, however, the original input matrix does not contain any structure. Therefore, our sparsity-aware algorithm does not reduce the remote memory request sizes (without using graph partitioning), compared to a sparsity-oblivious algorithm. After partitioning the eukarya dataset into 64 parts using METIS with estimated flops as the vertex weight, we achieve $2.05\times$ speedup excluding the permutation cost or $1.27\times$ speedup including the permutation cost. The METIS partitioning cost is 3.9 seconds for eukarya on a Perlmutter CPU node.

\begin{figure}[t]
    \centering
    \begin{minipage}{.8\linewidth}
    \includegraphics[width=\linewidth]{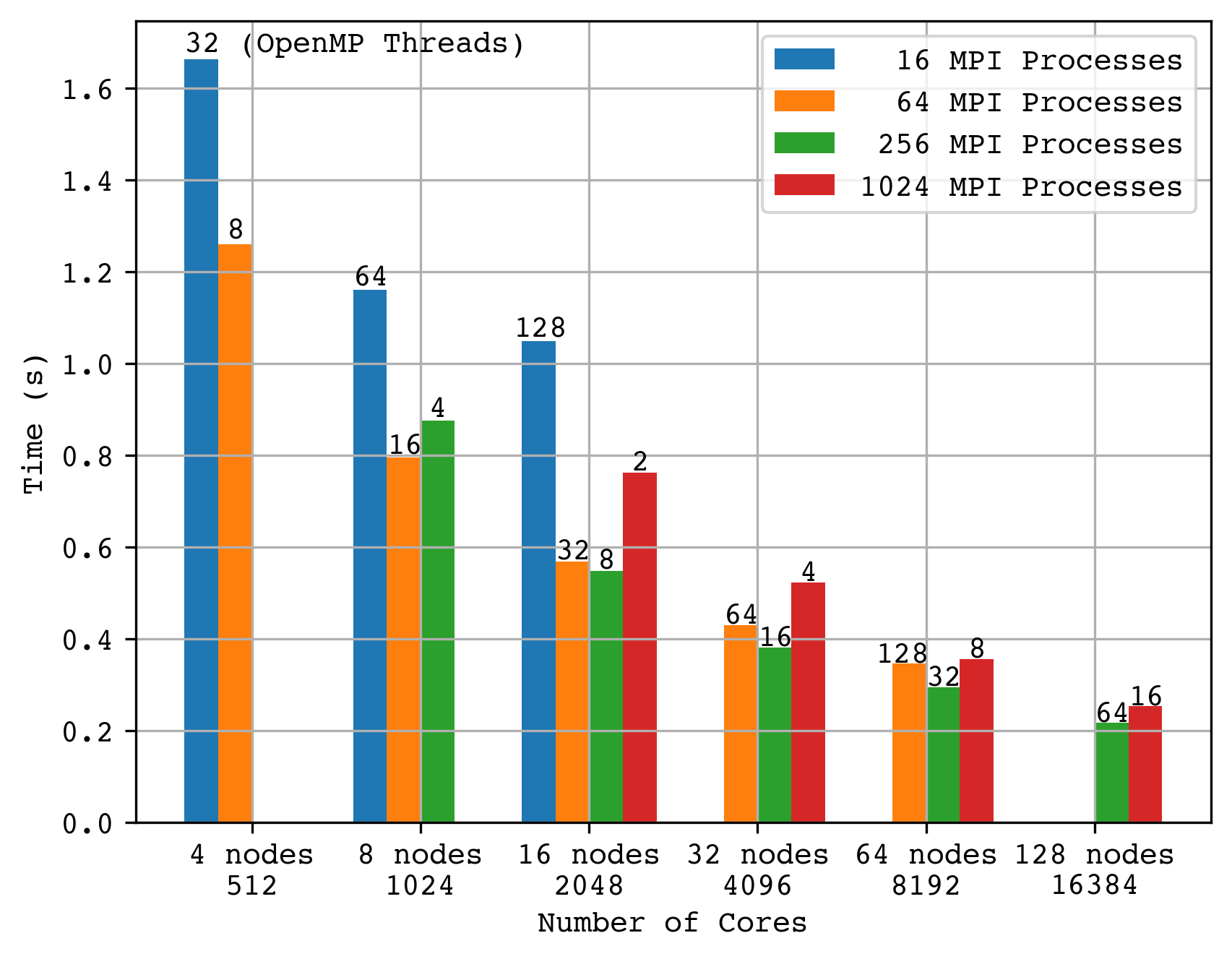}
    \caption{Performance comparison of different MPI + OpenMP configurations on hv15r dataset in squaring operation.}
    \label{fig:mpiompconfig_squaring_hv15r}
    \end{minipage}
    \begin{minipage}{\linewidth}
    \includegraphics[width=\linewidth]{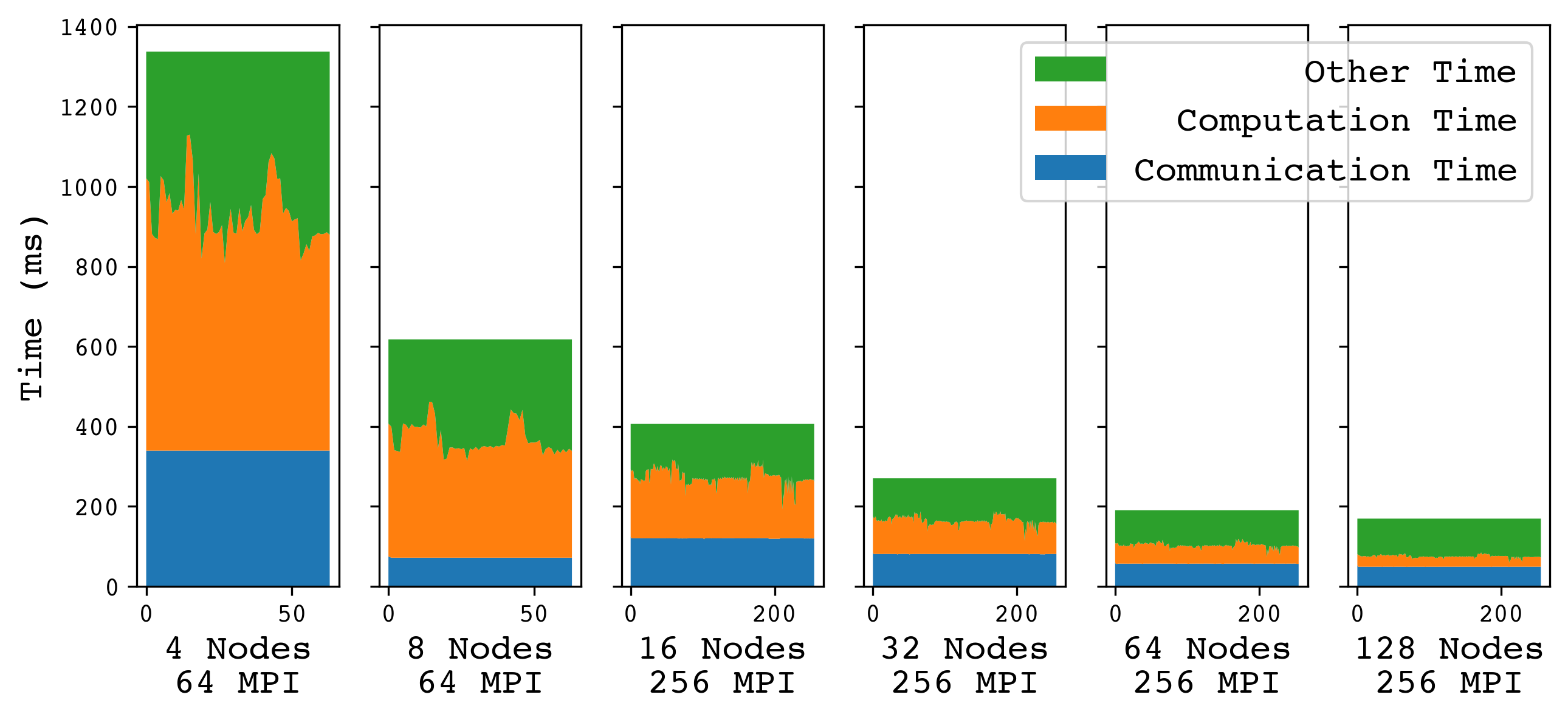}
    \caption{Breakdown of sparsity-aware 1D SpGEMM algorithm on hv15r dataset in squaring operation.}
    \label{fig:mpiomptimebreakdown_squaring_hv15r}    
    \end{minipage}
\end{figure}

\begin{figure*}[t]
    \includegraphics[width=\linewidth]{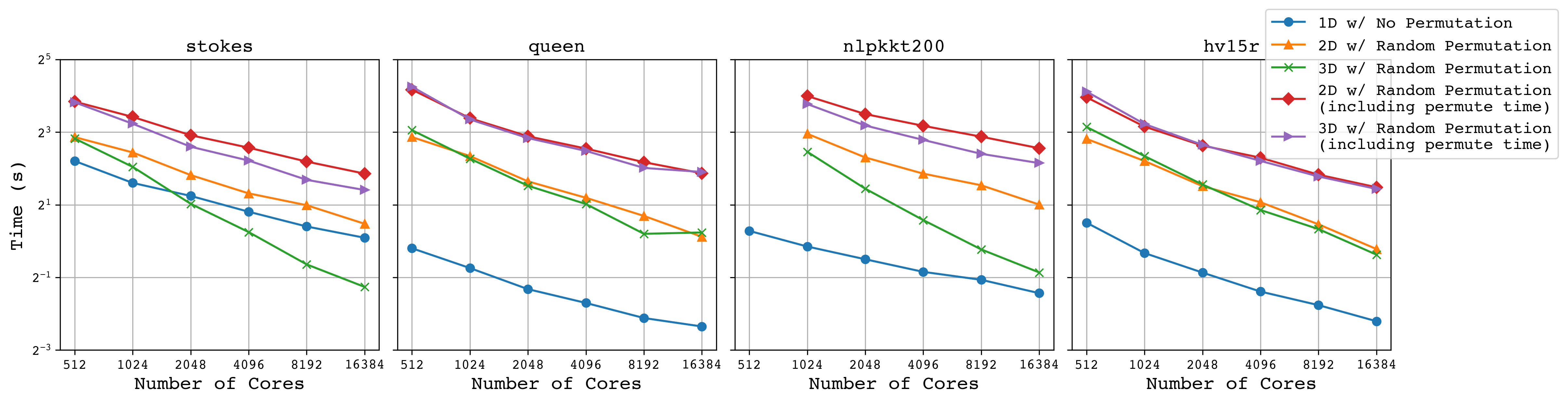}
    \caption{Strong scaling results of squaring operation using different SpGEMM algorithms on 4 datasets.}
    \label{fig:strongscaleAAall}
\end{figure*}

\subsubsection{RDMA Communication Analysis}

Figure~\ref{fig:eukhv15r_commana_AA} shows the communication volume reduction on hv15r and eukarya dataset in squaring operation using 4 nodes and 64 MPI processes, each with 8 OpenMP Threads. On hv15r dataset, we compare the random permutation with no permutation. On eukarya dataset we compare the random/no permutation with METIS permutation. In both cases, by using the correct permutation, we reduce the communication volume by around 96\%.

Figure~\ref{fig:hv15r_commana_AA} highlights the block fetch strategy on hv15r dataset in squaring operation. By using block fetching strategy, we significantly reduce the number of RDMA messages. We also improve the RDMA communication time due to reduced message size and latency cost.

\subsubsection{MPI + OpenMP Configuration Impact to Performance}
Figure~\ref{fig:mpiompconfig_squaring_hv15r} describes different MPI and OpenMP configurations when using sparsity-aware 1D SpGEMM algorithm on the hv15r dataset. In this evaluation, we study the performance impact of varying OpenMP threads and MPI processes for a given compute allocation. In other words, given $c$ cores, we are varying the number of threads ($t$) and processes ($p$) while keeping $c=p \cdot t$ constant.

We find that the algorithm prefers the intermediate configurations where the number of MPI processes ranges from 64 to 256. There are two reasons for this observation. First, in small number of MPI processes cases (such as the 16 MPI processes setting on 4, 8 and 16 nodes), the overhead of sequential code like copying the memory buffer damages the performance. Second, in the large number of MPI processes cases (such as he 1024 MPI processes setting on 32, 64 and 128 nodes), the communication time dominates the overall time to solution.

Figure~\ref{fig:mpiomptimebreakdown_squaring_hv15r} shows strong scaling results of hv15r dataset in squaring operation. We present the breakdown time of each MPI processes because we observe the load imbalance in the algorithm. Some amount of load imbalance is expected in the sparsity-aware 1D SpGEMM algorithm. However, such load imbalance is tamed in the large concurrency cases thus has less impact to the performance.

\subsubsection{Strong Scaling}

Figure~\ref{fig:strongscaleAAall} presents the strong scaling results across four datasets, alongside a comparison with other distributed SpGEMM algorithms. \new{In these strong scaling experiments}\Aydin{you mean, ``in these strong scaling experiments'', because, you did use permutations in other experiments certainly}, we did not apply any permutation for the sparsity-aware 1D SpGEMM algorithm across all datasets. For the 2D and 3D algorithms, random permutation was employed to achieve load balancing. We provide results for the 2D and 3D algorithms both with and without the inclusion of random permutation time. For the 3D algorithm, we explored all possible layer parameters and selected the optimal configuration for presentation.

The results clearly demonstrate that the sparsity-aware 1D SpGEMM algorithm exhibits strong scaling across all four datasets. Notably, when comparing strong scaling performance on the hv15r and queen datasets, our algorithm is an order of magnitude faster than the 2D and 3D algorithms, even when considering only the SpGEMM kernel time. For the stokes and nlpkkt200 datasets, our algorithm remains faster than the 2D and 3D algorithms when the permutation time is included.



\subsection{Multiplication with the Restriction Operator \mat{R}}

\begin{table}[ht]
    \centering
    \caption{Statistics information of \textbf{restriction operator} matrices. Each row of the restrction operator matrcies has exactly one non-zero element.}
    \begin{tabular}{l r r r r r}
    \toprule
    Dataset & $nrows(\mat{R})$ & $ncols(\mat{R})$ & $\nnz(\mat{R})$ \\
    \midrule
    queen\_4147 & 4147110 & 29972 & 4147110\\
    stokes     & 11449533 & 300473 & 11449533\\ 
    hv15r     & 2017169 & 7153 & 2017169 \\
    nlpkkt200 & 16240000 & 309097 & 16240000 \\
    \bottomrule
    \end{tabular}    
    \label{tab:ropsize}
\end{table}

\begin{figure}[ht]
\centering
\includegraphics[width=.6\linewidth]{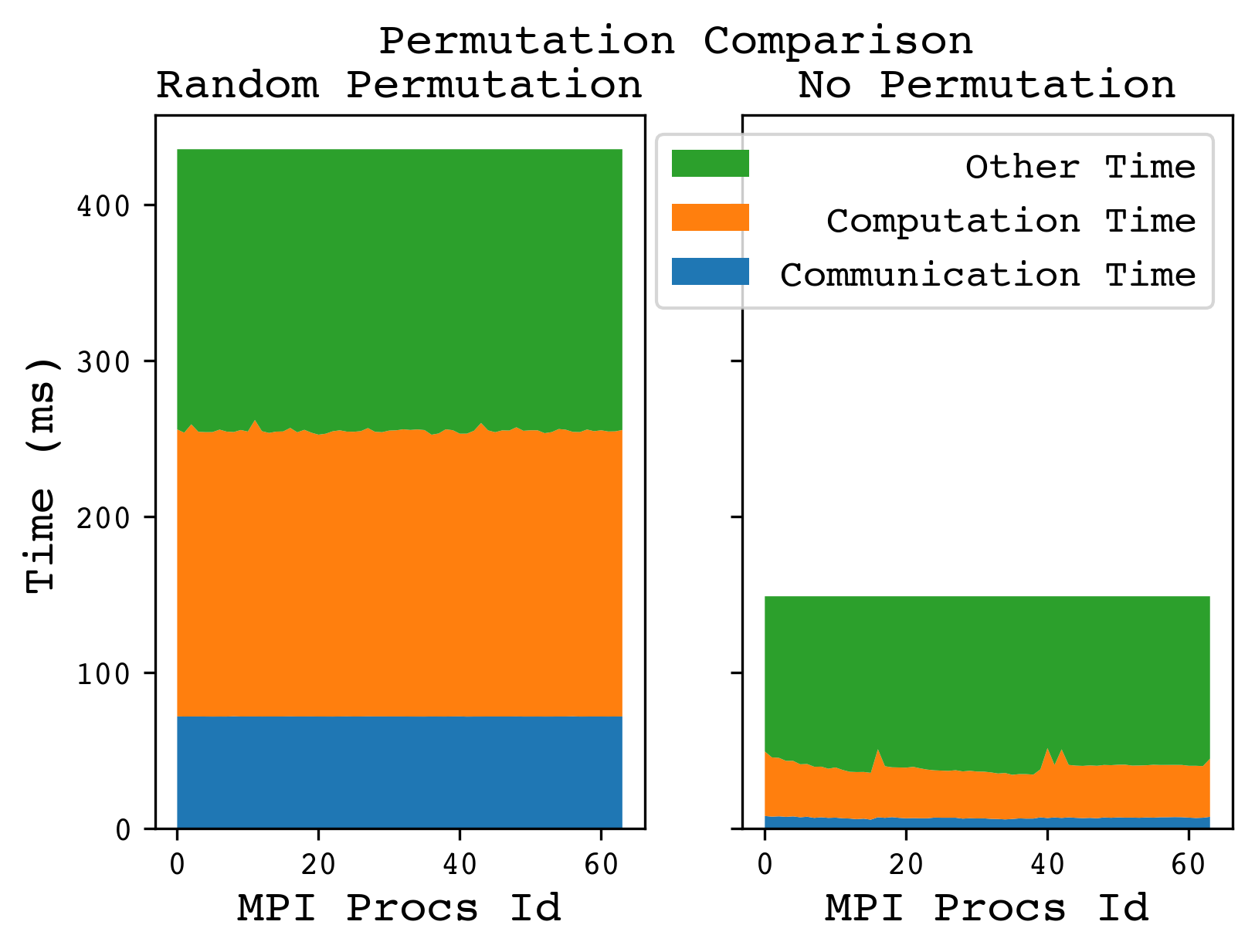}
\caption{Permutation comparison on $\mat{R}^T\mat{A}$ using queen dataset on 4 Perlmutter compute nodes.}
\label{fig:timebreakdown_RTA}
\end{figure}

\begin{figure*}[ht]
\centering
\includegraphics[width=\linewidth]{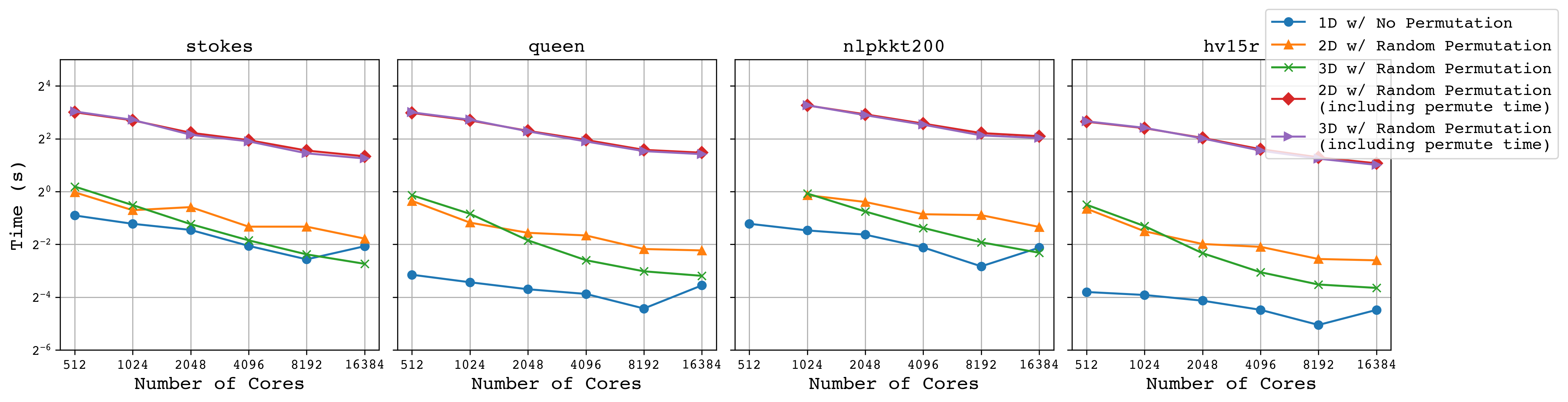}
\caption{Strong scaling results of $\mat{R}^T\mat{A}$ operation using different SpGEMM algorithms on 4 datasets.}
\label{fig:strongscalingRop}
\end{figure*}

\begin{figure}[ht]
\centering
\includegraphics[width=\linewidth]{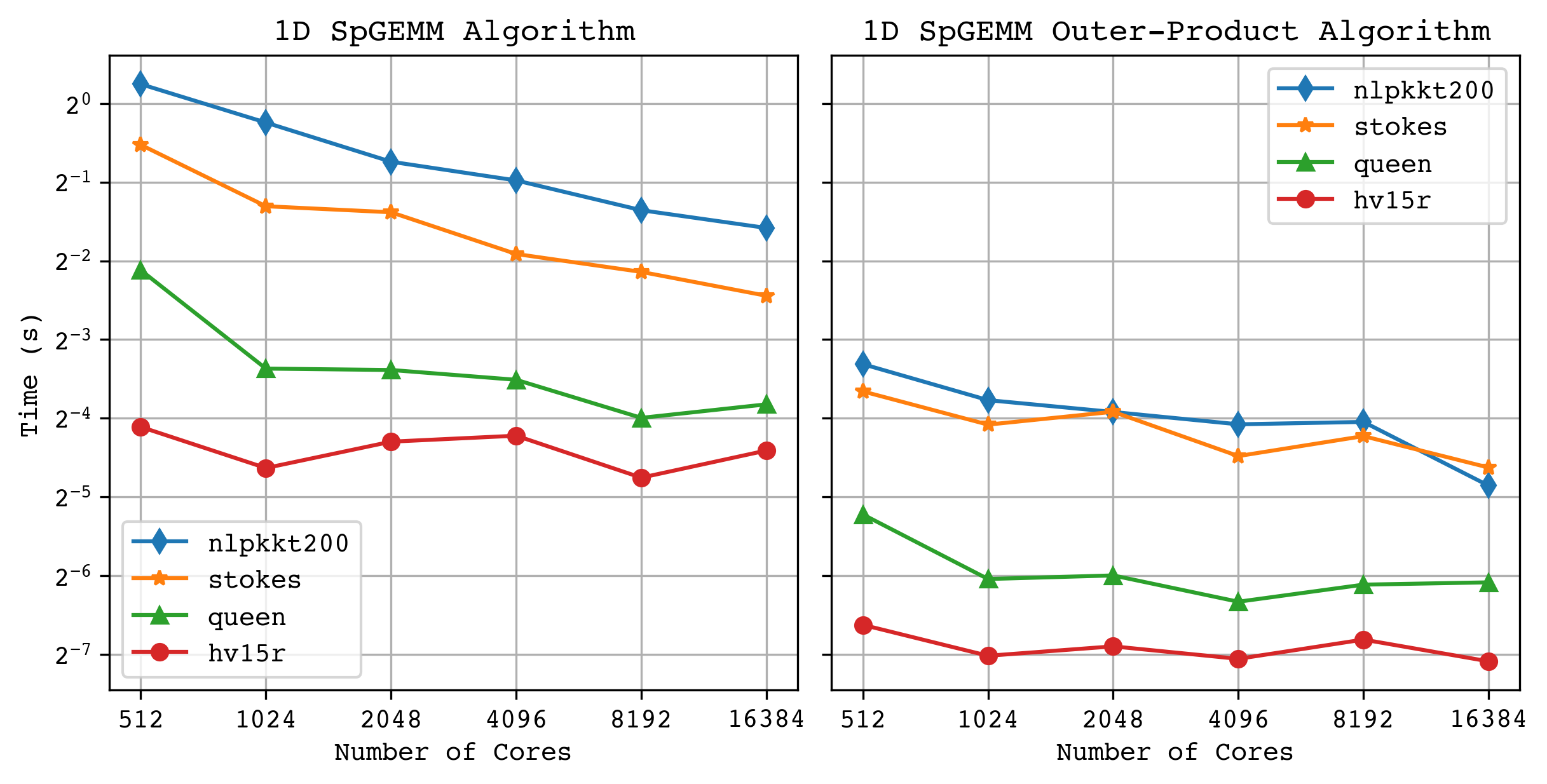}
\caption{Comparison of 1D SpGEMM algorithm and 1D SpGEMM outer-product algorithm on $(\mat{R}^T\mat{A})\mat{R}$.}
\label{fig:Rop_1dvs1dop}
\end{figure}

Table~\ref{tab:ropsize} describes the dimension of restriction operator and non-zero counts for each dataset. 
Figure~\ref{fig:timebreakdown_RTA} shows the time breakdown of per MPI processes on 4 nodes, 64 MPI processes each with 8 OpenMP threads on queen dataset in $\mat{R}^T\mat{A}$ operation. Compared to random permutation, using the original dataset  significantly reduces the communication time and local computation time.
It is worth noting that other time dominate the total time to solution. This is because the workload is not large enough and we have not seen load imbalance issue in this dataset. 

Figure~\ref{fig:strongscalingRop} shows the scaling results of sparsity-aware 1D SpGEMM algorithm on 4 datasets in $\mat{R}^T\mat{A}$ operation. The algorithm does not scale after 8192 cores. This is due to insufficient workload in the restriction operators. It also compares the scaling results of SpGEMM variants on queen dataset in restriction operators. We sum up times of $\mat{R}^T\mat{A}$ and $(\mat{R}^T\mat{A})\mat{R}$. Among the two of them, time of $\mat{R}^T\mat{A}$ is the dominant one. 1D SpGEMM variant is better than all other 2D, 3D algorithms. 

Figure~\ref{fig:Rop_1dvs1dop} presents the performance comparison of sparsity-aware 1D SpGEMM algorithm with the outer-product algorithm. The outer-product algorithm is better than sparsity-aware 1D SpGEMM algorithm for this use case.

\subsection{Betweenness Centrality}
\begin{figure}[ht]
\centering
\begin{minipage}{\linewidth}
\includegraphics[width=\linewidth]{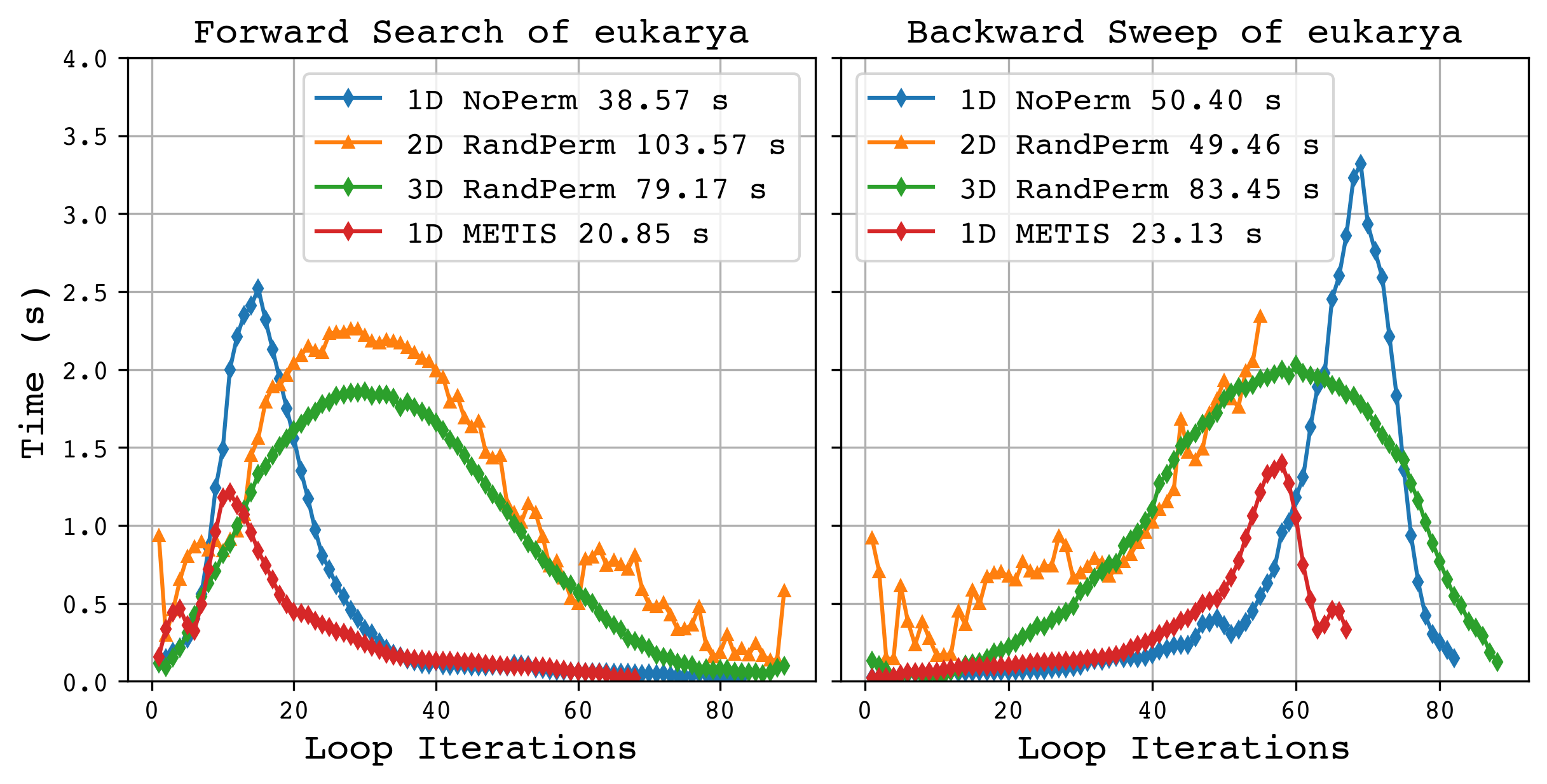}
\caption{Forward search and backward sweep of the first batch of eukarya dataset.}
\label{fig:bceuk}
\end{minipage}  
\begin{minipage}{\linewidth}
\includegraphics[width=\linewidth]{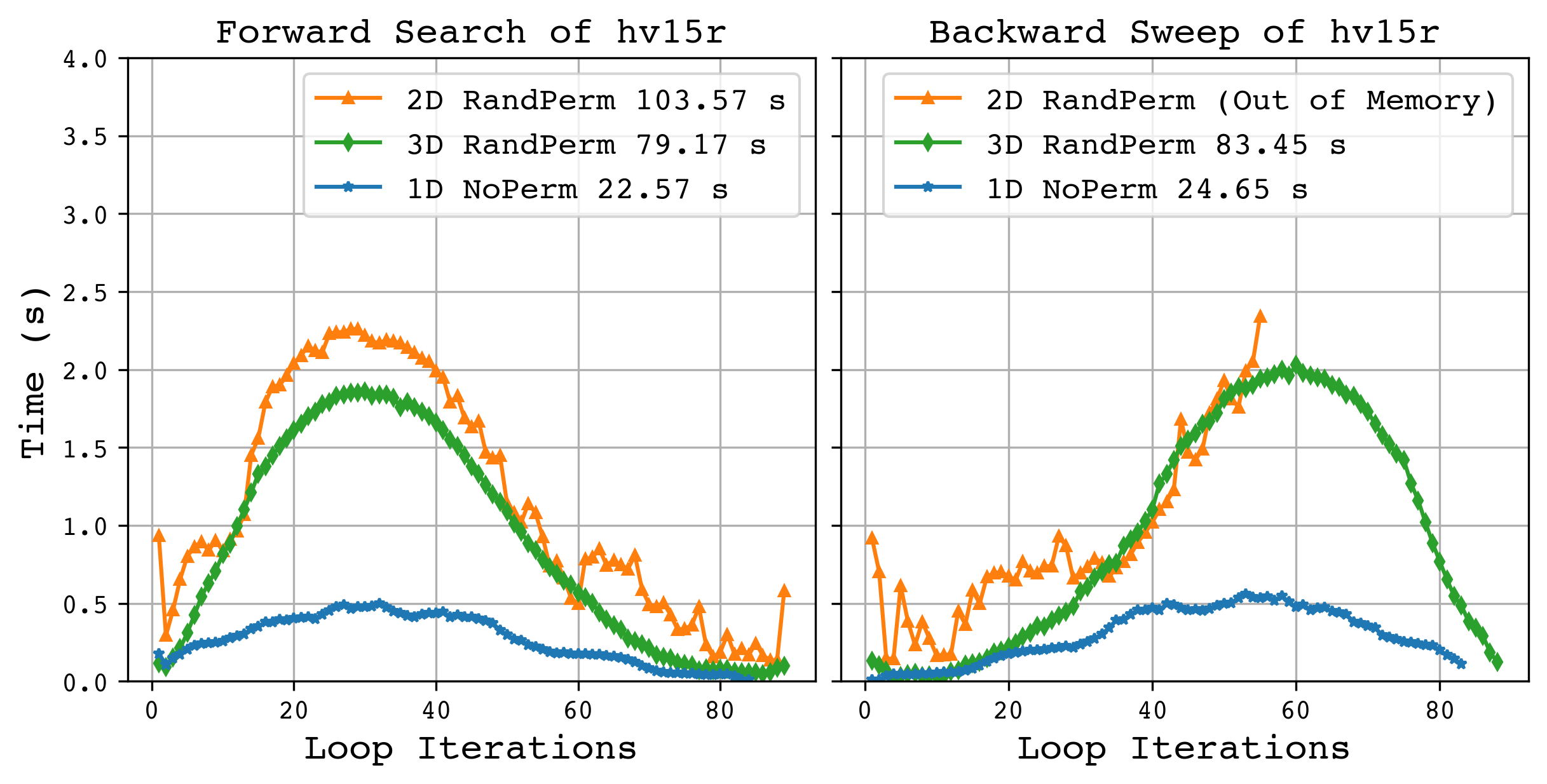}
\caption{Forward search and backward sweep of the first batch of hv15r dataset.}
\label{fig:bchv15r}
\end{minipage}
\end{figure}
In all betweenness centrality (BC) experiments, we do not report the partitioning overhead because partitioning is only performed once whereas SpGEMM is performed tens of thousands of times to compute a good approximation of BC scores. For example, using a 5\% sampling rate to compute approximate BC scores~\cite{bader2007approximating} on a 20M vertex graph would require 1M BFS searches. Even if we perform 1000 BFS searches in a batch using multi-source BFS, that still requires 1000 multi-source BFS executions, each execution requiring as many SpGEMM operations as the diameter of the graph. Therefore, partitioning time is negligible compared to the total time to solution in BC experiments. 

Our batch size is set to 4096. We benchmark only the forward search and backward sweep in the first batch of BC's algorithm. The reason is that the forward search and backward sweep is the most time consuming parts in each batch computation.
We choose two datasets as representatives from random permutation and METIS permutation. In Figure~\ref{fig:bceuk}, we present the SpGEMM time of each loop iterations in both forward search and backward sweep. We are using 4 nodes 64 MPI processes, each with 8 OpenMP threads. It is shown in the squaring section that eukarya  benefits from METIS permutation instead of random permutation. As is shown in the figure, when using METIS permutation, we are 1.74x faster than the second fast one -- the 3D algorithm.

In Figure~\ref{fig:bchv15r}, we present the results on hv15r dataset. we are using 8 nodes 64 MPI processes, each with 16 OpenMP threads. The 2D algorithm is running out of memory in backward sweep phase. Sparsity-aware 1D SpGEMM algorithm is also significantly faster than the 3D algorithm. We achieve 3.5x speedup compare to the state-of-the-art 3D algorithm.





\section{Discussion}
\label{sec:discussion}
\subsection{The criteria of applying graph partition on the original matrices}
To clarify the decision-making process regarding whether to use graph partitioning or random permutation, it is important to consider the nature of the dataset. The choice depends largely on the dataset's structure. If the dataset exhibits a high degree of existing structure, we recommend against using graph partitioning. Conversely, if the dataset is more randomly distributed, applying graph partitioning before the SpGEMM operation will benefit the SpGEMM operation.
More precisely, we recommend utilizing the ratio of communication volume to the size of the full matrix A (denoted as CV/memA) as a criterion for determining the application of graph partitioning methods. This parameter can be calculated prior to initiating actual RDMA communication, and the process is computationally lightweight. If the CV/memA of the original matrix exceeds a specified threshold (e.g., 30\%), it is advisable to apply graph partitioning to the dataset before proceeding with the computation of SpGEMM. For instance, the CV/memA depicted in Fig. 5(b) is 1.0, indicating that each MPI process must retrieve the entire matrix A to compute its local matrix C. It is generally recommended to test both scenarios and select the one that yields better performance.

\section{Conclusion}
\label{sec:conclusion}
We present a distributed-memory, sparsity-aware 1D SpGEMM algorithm in this paper. It utilizes the sparsity structure of the dataset and remote direct memory access (RDMA) technology to fetch only the data needed for each processor's local computation. This approach avoids the random permutation used in 2D and 3D sparsity-oblivious methods for load balancing, which is not optimal for 1D SpGEMM algorithms. Instead, it leverages the inherent sparsity pattern or METIS partitioning to distribute the workload effectively. To minimize the number of RDMA message calls, we employ a block-fetching strategy that groups remote data into blocks and fetches it using a limited number of RDMA messages. We also provide a detailed time breakdown analysis and RDMA call analysis of the sparsity-aware 1D SpGEMM algorithm.
The proposed algorithm demonstrates better performance than state-of-the-art 2D and 3D algorithms on datasets with specific sparsity patterns, such as in squaring, restriction operators, and betweenness centrality. Our 1D SpGEMM algorithm also scales well when the appropriate permutation method is chosen. We believe our 1D SpGEMM algorithm is a high-performance primitive that offers an alternative for SpGEMM-related applications. It is also easy to integrate with other scientific software like PETSc and Trilinos, as they both use a 1D process grid.

\section*{Acknowledgment}
This research was supported by the Office of Science of the DOE under Award Number DE-AC02-05CH11231. We used resources of
the NERSC supported by the Office of Science of the DOE under
Contract No. DE-AC02-05CH11231.

\bibliographystyle{plain}
\bibliography{main.bib}

\end{document}